\documentclass{mn2e}
\usepackage{epsfig} 
 
\renewcommand{\d}{{\rm d}}
\newcommand{\beq}{\begin{equation}}
\newcommand{\eeq}{\end{equation}}
\newcommand{\beqa}{\begin{eqnarray}} 
\newcommand{\eeqa}{\end{eqnarray}}
\newcommand{\bea}{\begin{array}} 
\newcommand{\ea}{\end{array}} 
\newcommand{\lag}{\langle}
\newcommand{\rag}{\rangle}
\newcommand{\Om}{\Omega_{\rm m}}

\newcommand{\wde}{w_{\rm de}}
\newcommand{\cP}{{\cal P}}
\newcommand{\Map}{M_{\rm ap}}

\newcommand{\inta}{\int_{-i\infty}^{+i\infty}}

\newcommand{\phiMap}{\varphi_{\Map}}
\newcommand{\phiM}{\varphi_M}
\newcommand{\Pb}{\overline{P}}
\newcommand{\Fb}{\overline{F}}
\newcommand{\Ub}{\overline{U}}
\newcommand{\Lb}{\overline{L}}

\title[Constraints from measurements of one-point distribution]
{Do measurements of the one-point distribution of aperture-mass improve 
constraints on cosmology ?}

\author[Munshi \& Valageas]
{Dipak Munshi$^{1,2}$, Patrick Valageas$^{3}$\\
$^{1}$Institute of Astronomy, Madingley Road,
Cambridge, CB3 OHA, United Kingdom\\
$^{2}$Astrophysics Group, Cavendish Laboratory, Madingley Road, 
Cambridge CB3 OHE, United Kingdom\\
$^{3}$Service de Physique Th\'eorique, 
CEA Saclay, 91191 Gif-sur-Yvette, France \\}

\begin{document}
\maketitle

\begin{abstract}

We study the possibility of using the entire probability distribution function 
(PDF) of the aperture mass $\Map$ and its related cumulative probability 
distribution function (CPDF) to obtain meaningful constraints on cosmological
parameters. Deriving completely analytic expressions for the associated 
covariance matrices, we construct the Fisher matrix and use it to estimate the 
accuracy with which various cosmological parameters can be recovered from 
future surveys using such statistics. This formalism also includes the effect
of various noises such as intrinsic ellipticity distribution of galaxies and 
finite survey size. The estimation errors are then compared 
with the ones derived from low order moments of the PDF (variance and skewness)
to check how efficiently the high $\Map$ tail can be used to constrain 
cosmological parameters such as $\Om$, $\sigma_8$ and dark energy equation of 
state $\wde$. We find that for future surveys such as JDEM the full PDF does 
not bring significant tightening of constraints on cosmology beyond what is 
already achievable by the joint use of second and third order moments.

\end{abstract}

 

\section{Introduction}

Weak lensing surveys can be used as a very efficient probe to constrain the 
background cosmology as well as contents of the Universe (see 
Munshi \& Valageas 2005 and references therein). Typically low order 
quantities such as two- and three-point functions or their reduced forms, 
variance and skewness, or the entire PDF are used to obtain cosmological 
constraints from observational data (Bernardeau et al. 2004). In weak lensing 
studies, observables such as $\kappa$, $\gamma$ or more 
commonly $\Map$ and their low order moments are extensively 
studied (e.g. Jarvis et al. (2004) for recent results for $\Map$). 
Use of $\Map$ has the additional advantage of being able to separate gravity 
induced shear signals (``Electric'' modes) from the (``Magnetic'') modes due 
to various systematics such as point spread function. 
The one-point probability distribution function of $\Map$ encodes information 
regarding non-Gaussianities at all orders (Bernardeau \& Valageas 2002) and 
thus can be useful to pinpoint background cosmological parameters by 
breaking degeneracies which appear at the level of two-point correlation 
functions (Schneider et al. 2002). We develop an analytical formalism to study 
the covariance of binned PDF and employ Fisher formalism techniques to study 
covariance of estimation error of cosmological parameters from realistic weak 
lensing surveys. The purpose of this study is twofold. Firstly we check the
estimation error associated 
with cosmological parameters while using binned PDF as the primary observable. 
Secondly we use these results to infer how much tightening of constraints, if 
any, can be achieved in general by using higher order informations regarding 
non-Gaussianities. This has been a topic of discussion in the recent years 
using the lowest order non-Gaussian statistics, e.g. bi-spectrum or related 
collapsed one-point skewness (Takada \& Jain 2002, Kilbinger \& Schneider 2004,
Munshi \& Valageas 2005). Thus our results based on PDF extend such studies 
to include all-orders of non-Gaussianity. This {\it letter} is organised as 
follows: in section~2 we discuss the analytical results concerning covariance 
structure of the PDF and CPDF. Borrowing results from Valageas et al. (2004)
we show how the Fisher matrix can be constructed from PDF and CPDF data. In 
section~3 the numerical results are provided. Section~4 is left for 
discussion of our results.

\section{Analytical Results}

\subsection{Covariance of PDF of aperture-mass $\Map$}

We recall in this section the analytical results presented in 
Valageas et al. (2005). The aperture-mass $\Map$ can be written as a function
of the tangential shear $\gamma_{\rm t}$ as (Kaiser et al. 1994; 
Schneider 1996):
\beq
\Map= \int \d{\vec \vartheta} \; Q_{\Map}({\vec \vartheta}) \; 
\gamma_{\rm t}({\vec \vartheta}) ,
\label{MapQ}
\eeq
with (using the same filter as Schneider 1996):
\beq
Q_{\Map}({\vec \vartheta}) = \frac{\Theta(\vartheta<\theta_s)}{\pi\theta_s^2}
\; 6  \; \left(\frac{\vartheta}{\theta_s}\right)^2 
\left(1-\frac{\vartheta^2}{\theta_s^2}\right)  .
\label{QMap}
\eeq
Then, in order to measure the 
aperture-mass $\Map$ within a single circular field of angular radius 
$\theta_s$, in which $N$ galaxies are observed at positions 
${\vec \vartheta}_j$ with tangential ellipticity $\epsilon_{{\rm t},j}$, we 
can use the estimator $M$ defined by:
\beq
M= \frac{\pi\theta_s^2}{N} \sum_{j=1}^N Q_{\Map}({\vec \vartheta}_j) \; 
\epsilon_{{\rm t},j} .
\label{M}
\eeq
In the case of weak lensing the observed complex ellipticity $\epsilon$ is 
related to the shear $\gamma$ by: $\epsilon=\gamma+\epsilon_*$,
where $\epsilon_*$ is the intrinsic ellipticity of the galaxy. Assuming that
the intrinsic ellipticities of different galaxies are uncorrelated random
Gaussian variables, the cumulant of order $p$ of $M$ is:
\beq
\lag M^p\rag_c = \lag \Map^p\rag_c \left( 1+\frac{\delta_{p,2}}{\rho}
\right) \;\;\; \mbox{with} \;\;\; \rho = \frac{5N\lag \Map^2\rag}
{3\sigma_*^2} ,
\label{Mapc}
\eeq
where $\delta_{p,2}$ is the Kronecker symbol and 
$\sigma_*^2=\lag\epsilon_*^2\rag$ is the dispersion of the intrinsic 
ellipticity of galaxies.
Since the intrinsic ellipticities are Gaussian and we neglected any 
cross-correlation with the density field they only contribute to the variance 
of the estimator $M$ (note that $M^2$ is a biased estimator of 
$\lag \Map^2\rag$ because of this additional term). The quantity $\rho$
 measures the relative importance of the galaxy intrinsic ellipticities in 
the signal. They can be neglected if $\rho \gg 1$. Any Gaussian white noise 
associated with the detector can also be incorporated into the expression 
(\ref{Mapc}) by adding a relevant correction to $\sigma_*^2$. Finally, from 
eq.(\ref{Mapc}) we obtain for the generating function $\phiM$ of the 
estimator $M$:
\beq
\phiM(y)= \frac{1+\rho}{\rho} \; \phiMap\left( \frac{\rho}{1+\rho} y \right) 
- \frac{1}{1+\rho} \frac{y^2}{2} ,
\label{phiM}
\eeq
where we defined as usual the characteristic function $\phiM$ of any random 
variable $M$ by the logarithm of the Laplace transform of its PDF $\cP(M)$:
\beq
\cP(M) = \inta \frac{\d y}{2\pi i \lag M^2\rag_c} \; 
e^{[M y -\phiM(y)]/\lag M^2\rag_c} .
\label{PM}
\eeq
Thus, the PDF of the estimator $\cP(M)$ is related to the PDF of the 
aperture-mass $\cP(\Map)$ through eq.(\ref{phiM}). Of course, for small 
$\rho$ we recover the Gaussian (i.e. $\phiM(y)=-y^2/2$) as $M$ is dominated by
the galaxy intrinsic ellipticities, whereas for large $\rho$ we recover 
$\phiMap(y)$ as $M$ is dominated by the weak lensing signal ($M \simeq \Map$).

Thus, each circular field of angular radius $\theta_s$ yields a particular
value for the quantity $M$ defined in eq.(\ref{M}). If the survey contains
$N_c$ such cells on the sky, we can estimate the PDF $\cP(M)$ through the
estimators $P_j$ or $F_j$ defined by:
\beq
P_j = \frac{1}{N_c \Delta_j} \sum_{n=1}^{N_c} {\bf 1}_j(n) , \;\;\; 
F_j = \Delta_j P_j = \frac{1}{N_c} \sum_{n=1}^{N_c} {\bf 1}_j(n) ,
\label{PjFj}
\eeq
where ${\bf 1}_j(n)$ is the characteristic function of the interval 
$I_j=[M_j,M_{j+1}[$ of width $\Delta_j$, applied to the value $M(n)$ of $M$ 
measured in the cell $n$:
\beq
{\bf 1}_j(n) = 1 \;\; \mbox{if} \;\; M_j \leq M(n) < M_{j+1}, \;\;
{\bf 1}_j(n) = 0 \;\; \mbox{otherwise} .
\label{Ij}
\eeq
Here we restrict ourselves to non-overlapping intervals $I_j$.
The estimators $F_j$ are slightly more convenient than the $P_j$ as they
do not explicitly involve the widths $\Delta_j$ which can vary with $j$ or
even be infinite (on each side of a binning of $[M_{\rm min},M_{\rm max}]$).
Then, from the sets $\{P_j\}$ or $\{F_j\}$ we obtain an histogram which 
provides an approximation to $\cP(M)$:
\beq
\lag F_j \rag = \Fb_j \;\;\; \mbox{with} \;\;\; \Fb_j = 
\int_{M_j}^{M_{j+1}} \d M \; \cP(M) .
\label{Fbj}
\eeq
Thus, for small enough $\Delta_j$ we have $\Pb_j=\Fb_j/\Delta_j \simeq 
\cP[(M_j+M_{_j+1})/2]$. Next, assuming that different cells are well
separated so as to be uncorrelated the covariance matrix $C_{ij}^{FF}$ of
estimators $\{F_j\}$ is simply:
\beq
C_{ij}^{FF} = \lag F_i F_j \rag - \lag F_i\rag \lag F_j\rag 
= \frac{\delta_{ij} \Fb_i - \Fb_i \Fb_j}{N_c} .
\label{CF}
\eeq
In addition to the $F_j$ which measure the PDF we have also considered the
estimators $U_j$ (upper) and $L_j$ (lower) associated with the CPDF:
\beq
U_j= \sum_{i \ge j} F_i, \;\;\; \Ub_j= \int_{M_j}^{M_{\rm max}} \d M \;\cP(M) ,
\label{Uj}
\eeq
\beq
L_j= \sum_{i < j} F_i, \;\;\; \Lb_j= \int_{M_{\rm min}}^{M_j} \d M \;\cP(M) .
\label{Lj}
\eeq
Their covariance matrices are simply:
\beq
C_{ij}^{UU} = \frac{\Ub_{{\rm max}(i,j)} - \Ub_i \Ub_j}{N_c} , \;\;\;
C_{ij}^{LL} = \frac{\Lb_{{\rm min}(i,j)} - \Lb_i \Lb_j}{N_c} .
\label{CUCL}
\eeq
The second (product) term is of rank one but the presence of the first term 
makes the matrices non-degenerate. It is also possible to construct joint data 
vectors but this will not be pursued here as they all carry the same 
information as the set $\{F_j\}$. Note that if the $N_{\cP}$ intervals run 
over $]-\infty,\infty[$ so that $\sum_j F_j=1$ the covariance matrices are 
degenerate as there are only $N_{\cP}-1$ independent variables. Thus, in 
practice we consider the set $\{F_j\}$ with $j=1,..,N_{\cP}-1$. In this case
we also have $\Lb_j+\Ub_j=1$ which yields $C_{ij}^{LL}= C_{ij}^{UU}$.

\subsection{Parameter estimation using the noisy PDF}

To obtain the estimation error for cosmological parameters $\Theta_{\alpha}$ 
measured from the observables $F_j,U_j$ or $L_j$, which we generically denote 
by $X_j$, we employ the Fisher matrix formalism commonly used in the 
litterature (e.g. Tegmark et al. 1997). The information regarding estimation 
errors $\lag\Delta\Theta_{\alpha}^2\rag^{1/2}$ and their
cross-correlations $\lag\Delta\Theta_{\alpha}\Delta\Theta_{\beta}\rag$ 
are encoded in the inverse of the Fisher matrix ${\cal F}_{\alpha\beta}$ which 
can be constructed from the covariance matrix $C$ of the data vector and its 
derivative w.r.t. cosmological parameters $\Theta_{\alpha}$:
\beq
{\cal F}_{\alpha\beta} =  \sum_{i,j} \frac{\d X_i}{\d\Theta_{\alpha}} 
\left[C^{-1}\right]_{ij} \frac{\d X_j}{\d\Theta_{\beta}} .
\eeq
Here we ignored the next order correction terms which involve the derivatives
of the covariance matrices (Munshi \& Valageas 2005) as they are much smaller. 
The cosmological parameters that we consider for extraction from the data 
are $\Om$, $\sigma_8$ and $\wde$. In this {\it letter} we concentrate on one 
angular scale $\theta_s=2'$. Use of other scales would not change much the 
constraints as nearby scales are highly correlated and non-gaussianities 
measured at very small or very large angular scales are too noisy 
(e.g. Munshi \& Valageas 2005).

\section{Numerical results}

\subsection{Covariance matrix and the derivative vector}

\begin{figure}
\begin{center}
\epsfxsize=4.12 cm \epsfysize=5.7 cm {\epsfbox[55 38 240 266]{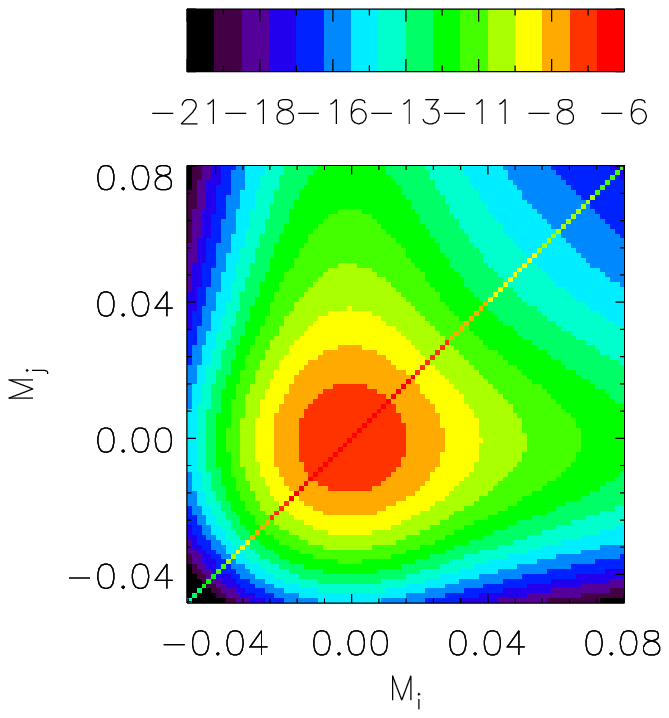}} 
\epsfxsize=4.12 cm \epsfysize=5.7 cm {\epsfbox[55 38 240 266]{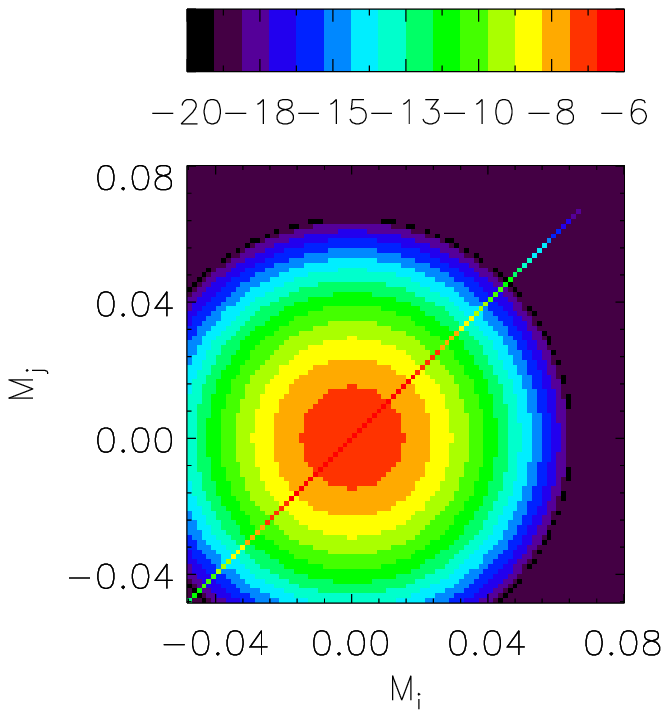}}
\end{center}
\caption{The covariance structure $C^{FF}_{ij}$ of the estimators $\{F_j\}$ 
of the non-linear PDF $\cP(M)$ (left panel) and its Gaussian counterpart 
(right panel). Non-linearities generate large correlations among various 
bins but the covariance remains diagonally dominated.}
\label{fig:pdfnl_cov}
\end{figure}

\begin{figure}
\begin{center}
\epsfxsize=4.12 cm \epsfysize=5.7 cm {\epsfbox[55 38 240 266]{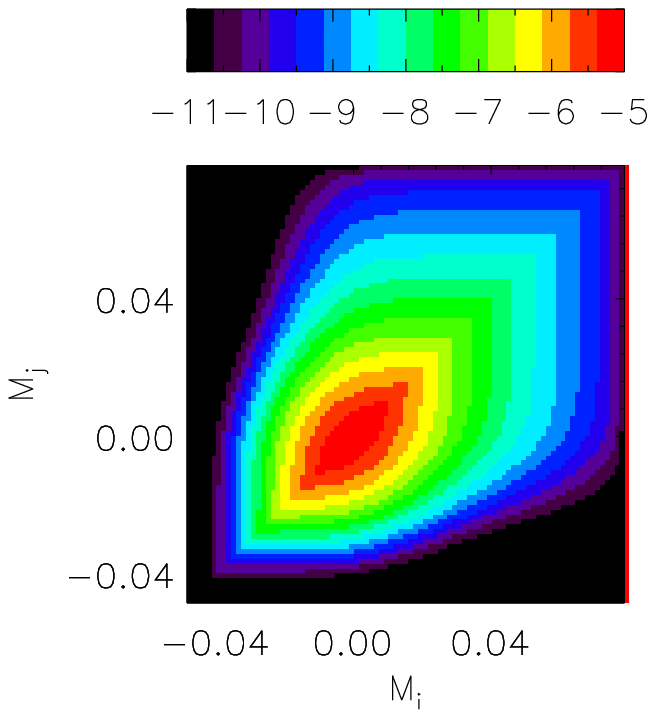}} 
\epsfxsize=4.12 cm \epsfysize=5.7 cm {\epsfbox[55 38 240 266]{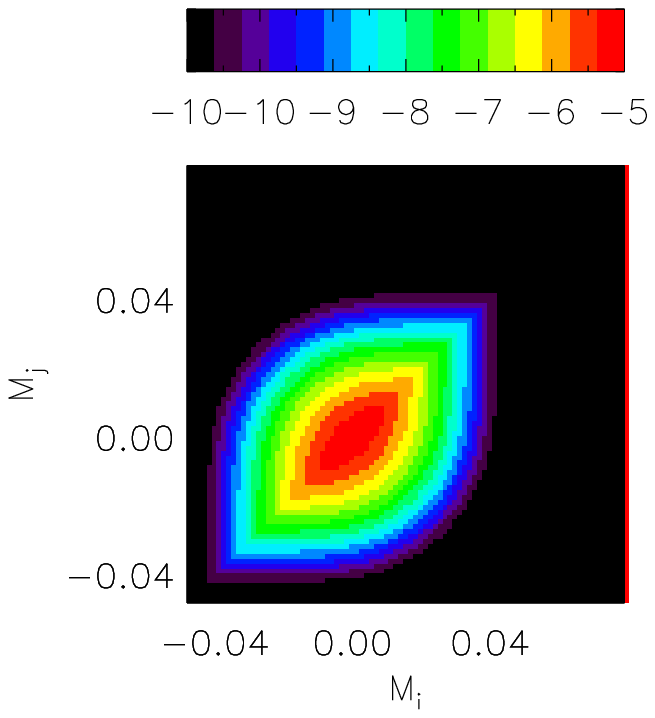}}
\end{center}
\caption{ The covariance structure of the estimators $\{L_j\}$ of the 
non-linear CPDF $\cP(<M)$ (left panel) and its Gaussian counterpart (right 
panel).}
\label{fig:cpdfnl_cov}
\end{figure}

\begin{figure}
\begin{center}
\epsfysize = 2.75truein {\epsfbox[40 3 450 350] {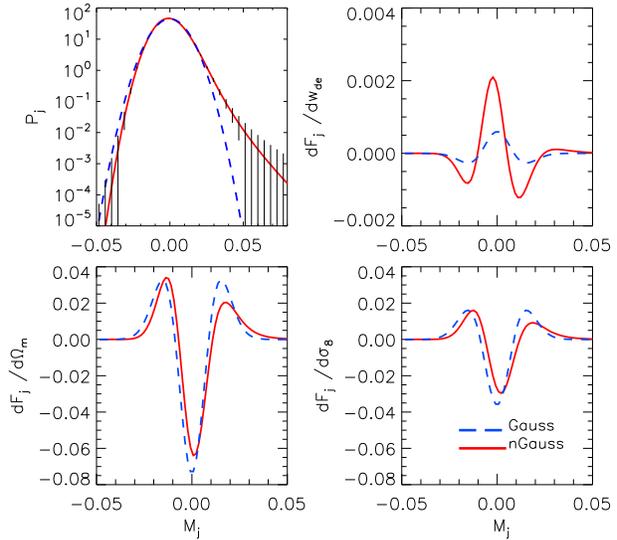}}
\end{center}
\caption{The PDF $\cP(M)$ (solid-line) and its Gaussian analog (dashed-line) 
are shown as a function of $M_j$ in the upper-left panel along with $1\sigma$ 
error bars. The corresponding derivatives of the associated estimators 
$\{F_j\}$ of these PDF are also displayed as a function of $M_j$ in other 
panels. We show the derivatives w.r.t. $\wde$ (upper-right panel), to $\Om$
(lower-left panel) and to $\sigma_8$ (lower-right panel).}
\label{fig:dpdf}
\end{figure}

For all numerical computations we have adopted a SNAP like survey strategy 
and a concordance $\Lambda$CDM cosmology as the fiducial model as in
Munshi \& Valageas (2005).

In Fig.~\ref{fig:pdfnl_cov} we plot the covariance matrix $C^{FF}_{ij}$ for 
$\{F_j\}$ using $N_{\cP}=91$ bins from $M=-0.05$ up to $M=0.08$. For 
comparison we plot in the right panel the covariance which would be obtained 
for a Gaussian PDF $\cP(M)$ with the same variance. For both cases there is 
a very pronounced diagonal dominance which makes them well-conditioned. Of 
course, the symmetry with respect to $M=0$ is broken when gravity-induced 
non-linearites are included. High $M$ tails are generated which follow the 
positive tails of $\cP(\Map)$ and $\cP(M)$ (Munshi et al. 2004, 2005) whereas 
a sharper decline appears at negative $M$, see also upper left panel of
Fig.~\ref{fig:dpdf}. We plot in 
Fig.~\ref{fig:cpdfnl_cov} the covariance matrix $C^{LL}_{ij}$ for the
CPDF (here $C^{LL}=C^{UU}$). Note that being integrated quantities the 
covariance matrix of $\{U_j\}$ or $\{L_j\}$ is more narrowly focussed along
the diagonal but we found that in practice the covariance matrix $C^{FF}$
is no more difficult to invert than $C^{LL}$.

The PDF $\cP(M)$ is shown in the upper-left panel of Fig.~\ref{fig:dpdf}
along with the derivatives of the data vector $\{F_j\}$ with respect to the 
three cosmological parameters $\wde$, $\Om$ and $\sigma_8$. The absolute 
value of the derivatives shows a well pronounced maxima near $M\simeq 0$ and 
two wings on each side. The symmetry is broken by inclusion of non-Gaussian 
effects as shown by the solid curves. Derivative vectors asymptotically reach
zero for high and low values of $M$, in bins where the PDF becomes small and
noise dominated and no useful cosmological information can be extracted. We 
can already see that the dependence
on $\wde$ is much smaller than the sensitivity to $\Om$ or $\sigma_8$ which
implies that errorbars for $\wde$ will be larger, as will be checked in
Fig.~\ref{fig:fishpdf} below.

\subsection{Parameter constraints using the PDF}

\begin{figure}
\begin{center}
\epsfxsize=3.8 cm \epsfysize=3.5 cm {\epsfbox[90 80 270 260]{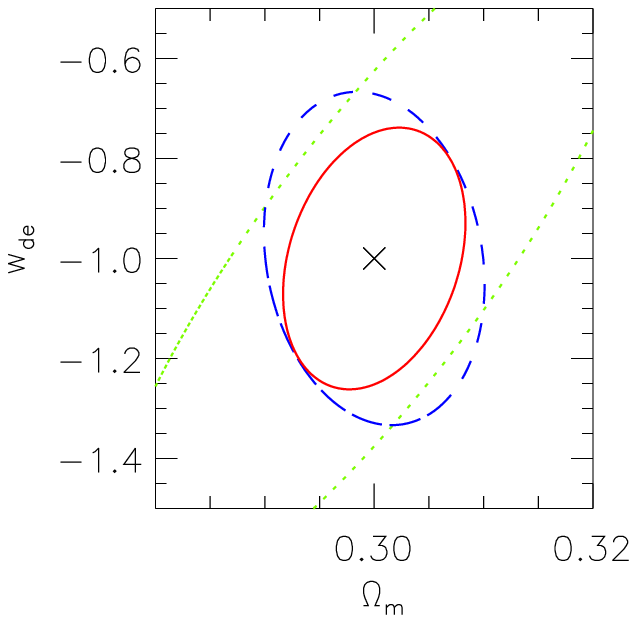}}
\epsfxsize=3.8 cm \epsfysize=3.5 cm {\epsfbox[90 80 270 260]{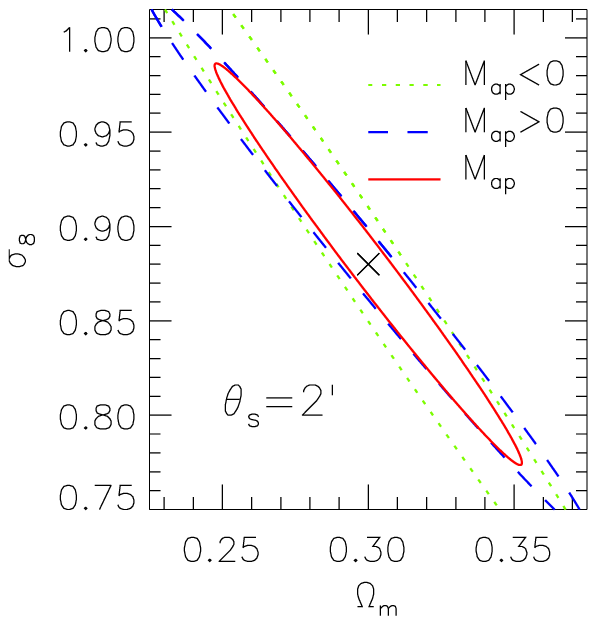}}
\end{center}
\caption{The contribution to the error-ellipses from various segments 
of the estimated PDF is shown for the cosmological parameter pairs
$\{\Om,\wde\}$ (left panel) and $\{\Om,\sigma_8\}$ (right panel). 
The dotted line corresponds to the $M<0$ sector of the PDF 
and the dashed line to the $M>0$ sector. The solid line corresponds to the 
entire range of $\{M_j\}$. All other parameters are assumed to be perfectly 
known. {\it All contours} in these panels as well as the next plots 
represent {\it $3\sigma$ contours}.}
\label{fig:fishpdf}
\end{figure}

\begin{figure}
\begin{center}
\epsfxsize=4.12 cm \epsfysize=4.95 cm {\epsfbox[90 56 270 266]{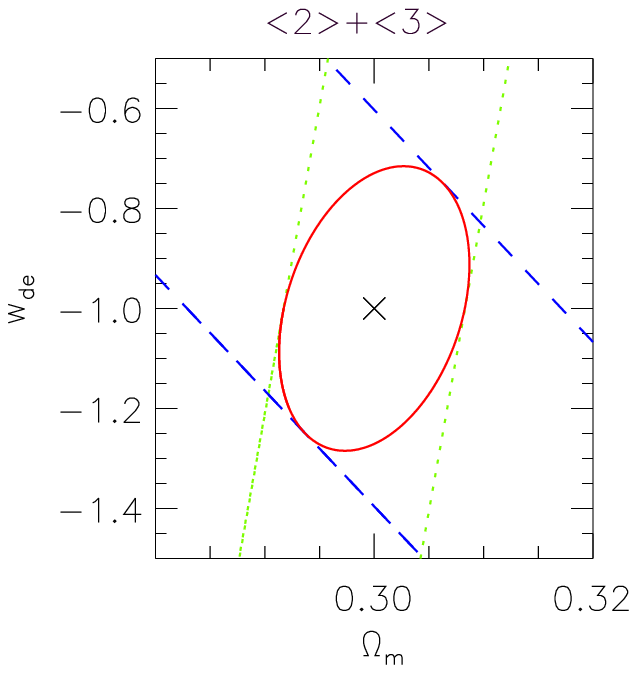}} 
\epsfxsize=4.12 cm \epsfysize=4.95 cm {\epsfbox[90 56 270 266]{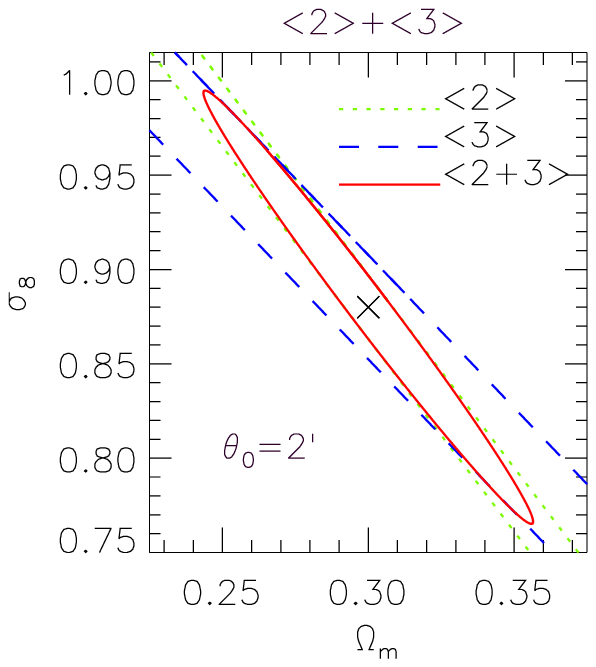}} 
\end{center}
\caption{The contribution to the error-ellipses from low-order moments are
presented for $\{\Om,\wde\}$ (left panel) and $\{\Om,\sigma_8\}$ 
(right panel). The dotted lines represent constraints from variance only
whereas the dashed lines represent constraints from skewness only. The solid 
lines show the joint constraints.}
\label{fig:fishmoments}
\end{figure}

From the the covariance and the derivative of the data vector $\{F_j\}$
we compute the Fisher matrix to study parameter degeneracies. We consider
two parameter pairs, $\{\Om,\sigma_8\}$ where parameters are negatively 
correlated and $\{\Om,\wde\}$ where they are positively correlated. We also 
use the two lowest order moments (variance and skewness) as data vectors to 
construct independent Fisher matrices. The latter are obtained from the
unbiased estimators $M_p$ defined as:
\beq
M_p= \frac{(\pi \theta_s^2)^p}{(N)_p} \sum_{(j_1,\dots,j_p)}^N 
Q_{\Map}({\vec \vartheta}_{j_1}) ... Q_{\Map}({\vec \vartheta}_{j_p}) \; 
\epsilon_{{\rm t}j_1} ... \epsilon_{{\rm t}j_p}
\label{Mp}
\eeq
where the sum runs over all sets of $p$ different galaxies among the $N$ 
galaxies enclosed in the angular radius $\theta_s$ and 
$(N)_p = N (N-1) ... (N-p+1)$. This ensures that 
$\lag M_p\rag=\lag\Map^p\rag$ (contrary to $\lag M^p\rag$). Thus the variance 
and skewness of $\Map$ are {\it not} estimated from the PDF $\cP(M)$.
The procedure is outlined in Munshi \& Valageas (2005) and will not be 
repeated here. As shown by the comparison of Figs.~\ref{fig:fishpdf} and
\ref{fig:fishmoments} we find that the constraints from the whole PDF are only
slightly tighter than the ones constructed from these two lowest order moments.
The similarity in shape is expected since our model for the non-linear PDF
is actually fully parameterized by the variance and skewness of the
underlying density field (see Munshi et al. 2004). Hence the PDF
and these two lowest order moments carry the same information within our
model. The good agreement of our analytical predictions with numerical
simulations (Munshi et al. 2004) shows that this must be true to a large 
extent for any realistic model. In a similar fashion, for the Gaussian 
approximation we noticed that the parameter constraints from only variance are 
in good agreement with the ones derived from PDF (which is then fully defined 
by its variance). However, it was not obvious that the size of the contour area
would be so close since the low-order estimators $\{M_2,M_3\}$ of 
eq.(\ref{Mp}) might have been more or less noisy than the data vector 
$\{F_j\}$. Indeed, the set $\{F_j\}$ contains many more points than the 
two-point data vector $\{M_2,M_3\}$ but it is based on the estimator $M$ which
is slightly more affected by the galaxy intrinsic ellipticities (e.g. 
although the variance of $M^2$ is equal to the variance of the estimator 
$M_2$ the variance of $M^3$ is somewhat larger than for $M_3$).
In particular, we checked that using as data vector
$\{\lag M^2\rag,\lag M^3\rag\}$ (i.e. the two lowest order moments of the 
PDF $\cP(M)$) yields confidence areas which are somewhat larger than for 
$\{M_2,M_3\}$.
Thus, it appears that although the full data vector $\{F_j\}$
is able to yield as good constraints as those obtained from $\{M_2,M_3\}$
the tails of the PDF are too noisy to provide significantly tighter 
constraints. Note that a drawback of $\cP(M)$ is that it is more difficult
to evaluate in surveys which contain many holes while low order moments may be
recovered by integrating the measured correlation functions.

\begin{figure}
\begin{center}
\epsfxsize=7.75 cm \epsfysize=4.5 cm {\epsfbox[65 60 432 242]{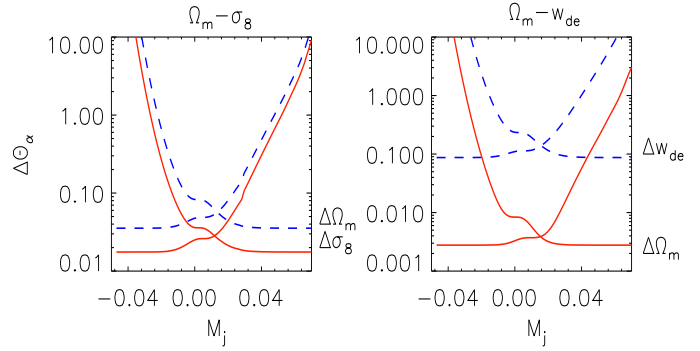}} 
\end{center}
\caption{The estimation error $\Delta\Theta_{\alpha}$ for two cosmological 
parameter pairs are shown, $\{\Om,\wde\}$ in right panel and 
$\{\Om,\sigma_8\}$ in left panel (hence this corresponds to $1\sigma$
errorbars). The lines 
which reach constant asymptotes for high positive values of $M_j$ are for the 
estimator $L_j$ and the ones which reach constant asymptotes for low negative
values of $M_j$ correspond to the estimator $U_j$. These asymptotes match 
exactly with the results based on the full estimator set $\{F_j\}$,
as CPDF and PDF carry equivalent information.}
\label{fig:fishcpdf_err}
\end{figure}

We have also investigated the amount of information which
can be extracted from the restriction of the PDF to $M<0$ (dotted lines
in Fig.~\ref{fig:fishpdf}) or $M>0$ (dashed lines). We can note that the
negative $M$ sector of the PDF generates ellipses which are more closely 
aligned with those from the variance analysis alone whereas the positive $M$ 
sector of the PDF produces ellipses which are more closely aligned with those 
from the skewness analysis. This is due to the fact that the skewness is mostly
sensitive to the large positive tail of $\cP(\Map)$ which is broader than the
negative tail (e.g. Munshi et al. 2004). On the other hand, the positive
sector $M>0$ alone already provides reasonable constraints on cosmology.
We also noticed that the 
cross-correlation terms in the covariance $C^{FF}_{ij}$ do not play any 
significant role and for a SNAP like survey can safely be ignored as far as 
estimation of cosmological parameters is concerned.
Since the CPDF data $U_j$ and $L_j$ are derived from the original data vector 
$\{F_j\}$ they cannot provide additional cosmological constraints. However, 
they can be used as a useful consistency check. We have checked that they give
exactly the same results as the PDF data vector $\{F_j\}$.

\begin{figure}
\begin{center}
\epsfxsize=7. cm \epsfysize=5. cm {\epsfbox[129 60 432 242]{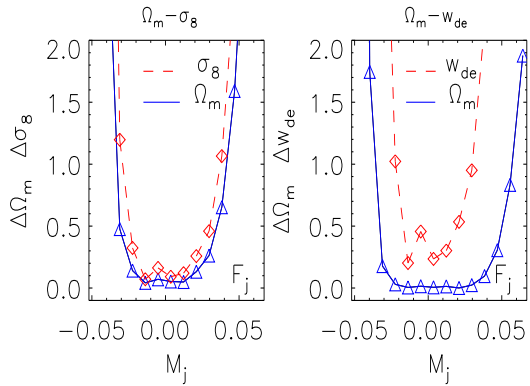}} 
\end{center}
\caption{The estimation error $\Delta\Theta_{\alpha} $ for two cosmological 
parameter pairs are shown for the estimators $F_j$,
$\{\Om,\wde\}$ in right panel and $\{\Om,\sigma_8\}$ in the left panel. A 
group of six neighbouring bins is used for each of these points 
to reduce the scatter in $\Delta\Theta_{\alpha}$.} 
\label{fig:fishpdf_err}
\end{figure}

Finally, we display the estimation error $\Delta\Theta_{\alpha}$ for the
pairs $\{\Om,\sigma_8\}$ and $\{\Om,\wde\}$ associated with different parts 
of the CPDF (Fig.~\ref{fig:fishcpdf_err}) or of the PDF 
(Fig.~\ref{fig:fishpdf_err}). The ``U-shape'' in Fig.~\ref{fig:fishpdf_err}
is consistent with the upper-left panel of Fig.~\ref{fig:dpdf}: the tails
of the PDF are too noisy to bring significant cosmological information.
On the other hand, the small decrease of $\Delta\Theta_{\alpha}$ near
$M_j \simeq 0.02$ for $\cP(<M)$ (solid lines in Fig.\ref{fig:fishcpdf_err})
is related to the secondary peak near $M_j \simeq 0.02$ in the derivatives 
shown in Fig.~\ref{fig:dpdf}. This feature contains significant cosmological
information which moreover goes beyond the mere variance as its shape differs
from the one obtained within a Gaussian approximation.

\section{Discussion}
\label{Discussion}

We have checked the possibility of using the entire $\Map$ PDF to
constrain cosmological parameters as opposed to the use of its lower order
moments which is more prevalent in the literature. We find that for a
SNAP like survey the PDF does not yield significantly tighter constrainsts 
than those
derived from the variance and skewness alone, despite the low-noise space 
based survey strategy. This also implies that higher order moments would not
contribute either to further tightening of error ellipses.
However, much larger surveys with high level of source galaxy
distributions might still benefit from measurement of kurtosis, but higher
number density requires increasing the depth of the survey which in turn
makes the PDF itself more Gaussian. Our results only take into account the 
volume averages of higher order correlation functions and did not propose 
to quantify non-Gaussianity beyond collapsed one point objects. It is
possible to extend our study by taking into account redshift binning
or tomography which we will report elsewhere, as well as combining several
angular scales. However, as seen in Munshi \& Valageas (2005)  
for a space based survey such as SNAP with reasonably good sky coverage 
and high number density of source galaxies most of the useful 
information can in effect be obtained by studying one particular 
angular scale (e.g. $\theta_s = 2'$ in case of SNAP, compare 
Fig.~\ref{fig:fishmoments} with Figs.14, 15 in Munshi \& Valageas 2005). 
Indeed, nearby scales are highly correlated and do not provide additional 
information whereas very large and very small angular scales are more affected 
by noise (due to the intrinsic ellipticity distribution of galaxies or the 
finite size of the survey).

The results provided here are for a monolithic survey strategy but real surveys
will have more complicated topology. However our results can provide 
clues as to what extent the surveys can probe non-Gaussianity to extract
meaningful constraints on cosmological parameters by using not only 
the first few lower order moments but the entire PDF.

We have not considered a Wiener filter based approach to reconstruct the PDF
from noisy data as suggested by Zhang \& Pen (2005) for convergence 
$\kappa$ maps but such an approach can very easily be implemented
by using the covariance matrices we have constructed. A complete Wiener 
filter based approach using compensated filter $\Map$ as presented here 
will be investigated elsewhere. In a different context shape of the 
non-Gaussian PDF was used by  Huffenberger \& Seljak (2005) to separate 
the kinetic-SZ contributions from primordial CMB. Similar approach can be
useful in separating gravity generated $E$ mode using $M_{ap}$ PDF 
from various systematics.Techniques presented here can also be applied to the
case of
weak lensing of diffuse background such as CMB (Kesden et al. 2002) and
high-redshift 21 centimeter radiation from neutral hydrogen
during the era of reionization (e.g. Cooray 2004).
Results of such analysis will be reported elsewhere.

\section*{acknowledgments}

DM acknowledges the support from PPARC of grant
RG28936. DM is pleased to thank members of CPAC
as well as M.~Kilbinger, L.~King, G.~Efstathiou and A.~Heavens
for many useful discussions.


\begin{thebibliography}{}
\bibitem{} Bernardeau F. et al., 2002,  Phys.Rept., 367, 1
\bibitem{} Bernardeau F. \& Valageas P.,  A\&A, 2000, 364, 1
\bibitem{} Cooray A., New Astron., 2004, 9, 173
\bibitem{} Huffenberger K.M., \& Seljak U., 2005, New Astron., 10
\bibitem{} Jarvis M.,  Bernstein G., Jain B., 2004, MNRAS, 352, 338
\bibitem{} Kesden M., Cooray A., Kamionkowski M., 2002, Phys.Rev. D66, 083007
\bibitem{} Kilbinger M. \& Schneider P. 2004, A\&A, 413, 465
\bibitem{} Munshi D. \& Valageas P. (2005) Phil.Trans.Roy.Soc.Lond. A363, 2675
\bibitem{} Munshi D. \& Valageas P. (2004), MNRAS, 354, 1146
\bibitem{} Munshi D., Valageas P. \& Barber A. (2004), MNRAS, 350, 77
\bibitem{} Schneider et al. (2002), A\&A, 396, 1
\bibitem{} Takada M. \& Jain B., (2004), MNRAS, 348, 897
\bibitem{} Tegmark M., Taylor A. \& Heavns A., (1997), Ap.J., 480, 22
\bibitem{} Valageas P., (2001), A\&A, 364, 1
\bibitem{} Valageas P., Munshi D., Barber A., (2005), MNRAS, 356, 386
\bibitem{} Zhang T., Pen U., (2005), ApJ in press. astro-ph/0503064


\end{thebibliography}
\end{document}